\documentclass[11pt]{article}
\usepackage{amsbsy}  % needed for bold Greek font
\usepackage{rotating}  % needed for landscape table

\begin{document}

\begin{center}{\Large\bf On a possible connection between Chandler wobble \\ and dark matter}
       \end{center}
\begin{center}{Oyanarte Portilho \\ 
{\it Instituto de F\'{\i}sica, Universidade de Bras\'{\i}lia, \\
70919-970 Bras\'{\i}lia--DF, Brazil}} \end{center}
\begin{center}{{\it e-mail:} {\tt portilho@fis.unb.br}} \end{center}

{\small Chandler wobble excitation and damping, one of the open problems in geophysics, is treated as
a consequence in part of the interaction between Earth and a hypothetical oblate ellipsoid
made of dark matter. The physical and geometrical parameters of such an ellipsoid and the 
interacting torque strength is calculated in such a way to reproduce the Chandler wobble
component of the polar motion in several epochs, available in the literature. It is also 
examined the consequences upon the geomagnetic field dynamo and generation of heat in the 
Earth outer core.}

\vspace{3mm}


\vspace{3mm}


\section{\bf Introduction}

A 305-day Earth free precession was predicted by Euler in the 18th century and was sought 
since then by astronomers in the form of small latitude variations. F. K\"ustner in 1888 and 
S. C. Chandler in 1891 detected such a motion but with a period of approximately 435 days.
Later it was shown by Love \cite{1love} and Larmor \cite{2larmor} that such a disagreement between theory and
observation is in part consequence of elastic deformation of the Earth which was not
considered by the Euler rigid body model (see Smith and Dahlen \cite{3smith} for further corrections
explanation). The so called Chandler wobble (CW) is actually one of the major components 
of the Earth polar motion, together with the annual wobble (AW) which has period of nearly
365 days. The composition of the two wobbles results in a motion with period of around 6.2
years due to the beat phenomenon. The ever increasing measurement precision, for which
techniques like very long baseline interferometry (VLBI), global positioning system
(GPS) and lunar laser ranging (LLR) are employed today, has lead to the conclusion that
both wobbles have varying amplitudes. For AW, this may be attributed to seasonal
displacement of atmospheric and water masses. On the other hand, excitation and
damping of the Chandler wobble has become a puzzle to investigators. Several
mechanisms have been proposed for its excitation like snow, hidrological,
atmospheric and oceanic mass displacements, and large earthquakes, with defenders and opponents
arguing for and against repeatedly and, in view of the large energy involved, the problem is
still an open question. In more recent papers, Gross \cite{4gross} attributes the excitation to ocean
bottom pressure variations and changes in ocean currents due to winds while Seitz {\it et al.} \cite{5seitz}
focus on combined effects due to atmospheric and oceanic causes.

Since it is believed that dark matter (DM) and dark energy composes around 96\% of the mass of the Universe we
investigate in this paper the possibility of DM to contribute, at least in part, for
the excitation and damping of Chandler wobble. The effect of DM upon regular matter has 
been considered mainly on galactic-scale although more recently some authors have studied smaller
scale effects. For instance, Froggatt and Nielsen \cite{6frni}) have investigated 
the possibility of star-scale effects due to small DM balls; Fr\`ere {\it et al.} \cite{fre}
studied the presence of a DM halo centered in the Sun and its influence upon the motion of  planets; numerical simulations due to Diemand {\it et al.} \cite{die} show the presence of DM clumps surviving near the solar circle; Adler \cite{adl} suggests that internal heat production in Neptune and hot-Jupiter exoplanets is due to accretion of planet-bound, not self-annihilating DM. Although the nature of the particles that form DM
is not known precisely (see for instance Khalil and Mu\~noz \cite{6khalil}) we assume that, as
usual, those particles do not interact with regular matter through electromagnetic force but
only gravitationally. Furthermore, we speculate that they can interact with each other in such
a way to set up an oblate (due to rotation) ellipsoid that might be present inside Earth, occupying the same
space without violating any physical law. It can be shown that such an ellipsoid, made of
self-interacting DM (otherwise such a structure would not be possible), would be
attached to Earth through a spring-like restoring force. In this line, excitation and damping
of CW could be, at least partially, consequence of energy exchange between the two bodies,
associated to the restoring torque between them. The model does not consider the other known
sources for the process nor the Earth internal structure, what are of course over
simplifications. However this becomes possible a simple starting point for the calculations, to
be sophisticated in a later step. H\"opfner \cite{7hop,8hop} has managed to isolate the CW and AW
components from available polar motion data. Our goal is to reproduce H\"opfner's results for
Chandler wobble in various epochs by adjusting the unknown physical and geometrical parameters
of the dark matter ellipsoid.
For this purpose, we describe both body motion by solving numerically the set of
differential equations that emerge from Euler equations and demonstrate that CW can be reproduced 
in such manner. This is explained in the next section while results and discussion are presented 
in section III.

\section{\bf The model}

We start by calculating the torque that arises when the polar principal axes 3 and 3$^{*}$ of
two oblate ellipsoids with isotropic densities interacting through the gravitational force are tilted
by an angle $\alpha$ (see Fig. 1). The gravitational potential at a point outside the internal
ellipsoid (see {\it e. g.} Stacey \cite{9stacey}) is given by
\begin{equation}
V(r,\theta)\approx -\frac{GM^{*}}{r}+\frac{G}{r^3}(C^{*}-A^{*})P_{2}(\cos\theta)  %eq (1)
\end{equation}
neglecting higher order terms. Here $M^{*}$, $A^{*}$ and $C^{*}$ are the mass, equatorial and
polar momenta of inertia of the internal ellipsoid, respectively, and 
$P_2(\cos\theta)=(3\cos^{2}\theta-1)/2$ is the Legendre polynomial of second order. The torque
over an infinitesimal mass $dm$ located at that point is then
\begin{equation}
d\tau=-dm\frac{\partial V}{\partial\theta}=\frac{3G(C^{*}-A^{*})}{2r^3}\sin(2\theta)\,dm.%eq. (2)
\end{equation}
and is directed orthogonal to the plane formed by the polar principal axis 3$^{*}$ and the
position $r$ of $dm$ (see Fig. 1). Considering the Earth tilted by an angle $\alpha$ around the
1-axis (Fig. 1) with respect to the internal ellipsoid it suffers therefore a torque given by
the integral
\begin{equation}
\tau=\int{\sin\phi\,d\tau}                         %eq. (3)
\end{equation}
directed towards the identical axes 1 and 1$^*$, as projected by the presence of $\sin\phi$
in this integral. The integration should in principle be performed over the entire Earth
volume. However considering the Earth density as isotropic, as it is done in the Preliminary
Reference Earth Model - PREM (Dziewonski and Anderson \cite{10dzie}), it can be shown that the spherical
volume internal to the Earth with radius the same as the polar radius does not contribute to
the integral. Therefore the integration over $r$ ranges from $R_p$, the polar radius, to the
radius of the geoid at a given direction. On the other hand, it is well known (see, for
instance, Stacey \cite{9stacey}) that the equation of the geoid is written in the first order
approximation in terms of the co-latitude $\theta'$ as
\begin{equation}
r_g\approx a(1-f\cos^{2}\theta'),       %eq. (4)
\end{equation}
where $a$ is the equatorial radius and $f=1-R_p/a$ is the flattening. Let us write
above equation for the case where the geoid is tilted by an angle $\alpha$ around the 1-axis
(Fig. 1), which yields
\begin{equation}
r_g\approx a\left[1-f(\cos\alpha\cos\theta-\sin\alpha\sin\theta\sin\phi)^2\right], %eq. (5)
\end{equation}
where $\phi$, together with co-latitude $\theta$, is part of the spherical coordinates in 
the 1$^*$2$^*$3$^*$ reference frame. Therefore the torque given by equation (3) becomes
\begin{eqnarray}
\tau & = & \frac{3}{2}\rho G(C^*-A^*)\int^{\pi}_{0} d\theta\,\sin(2\theta)\sin\theta \nonumber \\
   && \times \int^{2\pi}_{0} d\phi\,\sin\phi \ln \left\{ \frac{a}{R_p} \left[ 1-f
    (\cos\alpha\cos\theta-\sin\alpha\sin\theta\sin\phi)^2\right] \right\}, %eq. (6)
\end{eqnarray}
where $\rho$ is the average Earth mass density in that region, taken as 2600 kg/${\rm m}^3$ \cite{10dzie}.
Considering that $f=1/298.257<<1$, we can use the approximation $\ln (1+x) \approx x$, obtaining
\begin{equation}
\tau=c_1 \sin(2\alpha),                                                    %eq. (7)
\end{equation}
with $c_1 = \frac{4\pi}{5}\rho G(C^*-A^*)f$. Notice that this is a restoring torque when $\alpha$ is 
around 90$^{\circ}$, which gives the stable angular position if no initial angular momenta were involved.

We describe the Earth-ellipsoid motion by the Euler equations
\begin{eqnarray}
& & A \dot{\omega}_1 +(C-A)\omega_2\omega_3=\tau_1 \\   %eq. (8)
& & A \dot{\omega}_2 -(C-A)\omega_1\omega_3=\tau_2 \\   %eq. (9)
& & C \dot{\omega}_3=\tau_3=0 \ \ \Rightarrow \ \ \omega_3={\rm constant} \\ %eq. (10)
& & A^* \dot{\omega}^*_{1^*} +(C^*-A^*)\omega^*_{2^*}\omega^*_{3^*}=\tau_1^* \\  %eq. (11)
& & A^* \dot{\omega}^*_{2^*} -(C^*-A^*)\omega^*_{1^*}\omega^*_{3^*}=\tau_2^* \\  %eq. (12)
& & C^* \dot{\omega}^*_{3^*}=\tau_3^*=0 \ \ \Rightarrow \ \ \omega^*_{3^*}={\rm constant} %eq. (13)
\end{eqnarray}
where $A$ and $C$ are the equatorial and polar momentum of inertia of Earth, which is supposed to be 
axially symmetric ($B = A$), $\omega_1$, $\omega_2$, $\omega_3$ are the components of the Earth 
angular velocity $\boldsymbol{\omega}$ in the 123 reference frame, which is attached to Earth 
such that the 3-axis 
coincides with the polar principal axis, {\it i. e.}, the symmetry axis of the oblate ellipsoid, 1- 
and 2-axis are equatorial principal axes and $\tau_1$, $\tau_2$ 
are components of the torque suffered by Earth. Similar internal ellipsoid quantities are denoted by an 
asterisk. Notice that $\tau_3$ and $\tau_3^*$ vanish, what implies that $\omega_3$ and 
$\omega^*_{3^*}$ are constant. Notice also that we are not considering external torques, mainly 
exerted by the Sun and the Moon, which produce the precession of the equinoxes and is of no interest 
here. On the other hand, when $\tau_1 = \tau_2 = \tau_1^* = \tau_2^* = 0$ equations 
(8)-(13) lead to the well known Eulerian free precession motion.

We write the angular velocities in terms of the Euler angles $\theta$, $\phi$, $\psi$ and their 
derivatives in time, defined with respect to a fixed reference frame with axes X, Y, and Z (see Fig. 2). 
This frame is constructed such that the total angular momentum of the Earth-ellipsoid system, which 
is a constant of the motion, coincides with the Z-axis. The relations are the following
\begin{eqnarray}
  \omega_1 & = & \dot{\theta}\cos\psi +\dot{\phi}\sin\theta\sin\psi    \\    %eq. (14)
  \omega_2 & = & -\dot{\theta}\sin\psi +\dot{\phi}\sin\theta\cos\psi   \\     %eq. (15)
  \omega_3 & = & \dot{\phi}\cos\theta +\dot{\psi}=
       {\rm constant} \ \ \ \Rightarrow \ \ \  \dot{\psi} = \omega_3 - \dot{\phi}\cos\theta.  %eq. (16)
\end{eqnarray}
Similar relations for $\omega^*_{1^*}$, $\omega^*_{2^*}$, $\omega^*_{3^*}$ can be written in terms of 
the Euler angles $\theta^*$, $\phi^*$, $\psi^*$, referred also with respect to the XYZ fixed system, 
and their derivatives in time. Notice that equation (16) serves for the purpose of eliminating 
$\dot{\psi}$, while the same elimination can be done for $\dot{\psi^*}$. The components of the torque
$\tau$ over the Earth can also be expressed in terms of the Euler angles
\begin{eqnarray}
  \tau_1 & = & 2c_1\cos\alpha\left[-\sin\theta\cos\psi\cos\theta^*+(\cos\theta\cos\phi\cos\psi-
       \sin\phi\sin\psi)\sin\theta^*\cos\phi^*\right. \nonumber \\
   && \left. \mbox{}+(\cos\theta\sin\phi\cos\psi+\cos\phi\sin\psi)\sin\theta^*\sin\phi^* \right] \\ %eq. (17)
   \tau_2 & = & 2c_1\cos\alpha[\sin\theta\sin\psi\cos\theta^*-(\cos\theta\cos\phi\sin\psi+
       \sin\phi\cos\psi)\sin\theta^*\cos\phi^* \nonumber \\
   && \mbox{}+(-\cos\theta\sin\phi\sin\psi+\cos\phi\cos\psi)\sin\theta^*\sin\phi^* ].  %eq. (18)
\end{eqnarray}
The expressions for the components $\tau_1^*$ and $\tau_2^*$ of the torque $\boldsymbol{\tau}^*$ over the 
internal ellipsoid are like above, provided the exchanges $\theta \leftrightarrow \theta^*$, 
$\phi \leftrightarrow \phi^*$, $\psi \leftrightarrow \psi^*$ are made. In the same way, we also 
need the expression for $\cos\alpha$
\begin{equation}
  \cos\alpha=\sin\theta\sin\theta^*\cos(\phi-\phi^*)+\cos\theta\cos\theta^* .   %eq. (19)
\end{equation}
After substituting equations (14)-(18) into equations (8), (9), (11), (12) and considering that 
we are dealing with real quantities we can eliminate $\psi$, $\psi^*$, $\dot{\psi}$,  
$\dot{\psi}^*$, getting the following set of four second-order differential equations
\begin{eqnarray}
 & & \hspace{-1.6cm} \ddot{\phi}\sin\theta-\frac{C}{A}\omega_3\dot{\theta}+2\dot{\theta}\dot{\phi}\cos\theta-
     2\frac{c_1}{A}\cos\alpha\sin\theta^*\sin(\phi^*-\phi)=0 \\ %eq. (20)
 & & \hspace{-1.6cm} \ddot{\theta}-\dot{\phi}^2\sin\theta\cos\theta+\frac{C}{A}\omega_3\dot{\phi}\sin\theta
     +2\frac{c_1}{A}\cos\alpha\left[\sin\theta\cos\theta^*-\cos\theta\sin\theta^*
     \cos(\phi-\phi^*)\right]=0                               \\ %eq. (21)
 & & \hspace{-1.6cm} \ddot{\phi}^*\sin\theta^*-\frac{C^*}{A^*}\omega^*_{3^*}\dot{\theta}^*+2\dot{\theta}^*
      \dot{\phi}^*\cos\theta^*-2\frac{c_1}{A^*}\cos\alpha\sin\theta\sin(\phi-\phi^*)=0  \\ %eq. (22)
 & & \hspace{-1.6cm} \ddot{\theta}^*-\dot{\phi}^{*2}\sin\theta^*\cos\theta^*+\frac{C^*}{A^*}
     \omega^*_{3^*}\dot{\phi}^*\sin\theta^*
     +2\frac{c_1}{A^*}\cos\alpha\left[\sin\theta^*\cos\theta \right. \nonumber \\
 & &  \hspace{7.2cm} \left. \mbox{}-\cos\theta^*\sin\theta\cos(\phi-\phi^*)\right]=0 .            %eq. (23)
\end{eqnarray}

As a test for these equations we consider the case where $c_1 = 0$, what means that the 
ellipsoids are not interacting and therefore free precession occurs. It is known (see {\it e. g.} 
Symon \cite{11symon}) in this case that for each ellipsoid both polar principal symmetry axis (3-axis) 
and the rotation axis, which contains the angular velocity vector $\boldsymbol{\omega}$, 
have precession around the conserved angular momentum vector {\bf L} forming the so 
called body cone and space cone, with semi-angles $\alpha_b$ and $\alpha_s$, respectively 
(see Fig. 3). Notice that the 3-axis, $\boldsymbol{\omega}$ and {\bf L} are always in the 
same plane. The constant angular velocity of such a precession around fixed {\bf L} is 
\begin{equation}
  \dot{\phi}'=\beta\frac{\sin\alpha_b}{\sin\alpha_s}\omega_3=\frac{C}{A\cos\theta'}\omega_3 %eq. (24)
\end{equation}
with
\begin{equation}
  {\theta}'=\alpha_b-\alpha_s                    %eq. (25)
\end{equation}
being the constant angle that gives the orientation of the principal 3-axis with respect 
to fixed {\bf L}, and 
\begin{eqnarray}
  & & \tan\alpha_b = \frac{\sqrt{\omega_1^2+\omega_2^2}}{\omega_3}  \\     %eq. (26)
  & & \cos\alpha_s = \frac{1+\beta\cos^2\alpha_b}{\sqrt{\rule{0mm}{3mm}1+(2+\beta)\beta
         \cos^2\alpha_b}}                                          \\ %eq. (27)
  & & \beta = \frac{C}{A}-1.                    %eq. (28)
\end{eqnarray}
However, with two ellipsoids, we have a more general situation where the fixed reference 
frame Z-axis is taken to be in the direction of the total angular momentum ${\bf L}_t$, 
which is conserved if external torques are negligible, and not of the Earth angular 
momentum {\bf L}. Therefore we define a reference frame xyz, such that z is in the direction 
of {\bf L}, oriented with respect to XYZ through the Euler angles $\delta_\theta$, 
$\delta_\phi$, $\delta_\psi$ (see Fig. 4). While the principal 3-axis has precession 
around z-axis with constant angular velocity $\dot{\phi}'$ and angular amplitude 
$\theta'$, the 3-axis has motion described in terms of the Euler angles 
$\theta$, $\phi$ with respect to the frame XYZ as 
\begin{eqnarray}
 && \hspace{-1cm} \cos\theta=\sin\theta'\left[\sin(\dot{\phi}t+\epsilon)\sin\delta_\theta\sin\delta_\psi
   -\cos(\dot{\phi}t+\epsilon)\sin\delta_\theta\cos\delta_\psi\right]+
   \cos\theta'\cos\delta_\theta \\                 %eq. (29)
 && \hspace{-1cm} \tan\phi=-\frac{\sin\theta'\left[B_1\sin(\dot{\phi}t+\epsilon)+B_2\cos(\dot{\phi}t+
   \epsilon)\right]+\cos\theta'\sin\delta_\theta\sin\delta_\phi}
   {\sin\theta'
   \left[B_3\sin(\dot{\phi}t+\epsilon)+B_4\cos(\dot{\phi}t+\epsilon)\right]-
   \cos\theta'\sin\delta_\theta\cos\delta_\phi}    %eq. (30)
\end{eqnarray}
with $\epsilon$ being an arbitrary phase and
\begin{eqnarray}
  B_1 & = & \cos\delta_\phi\cos\delta_\psi-\cos\delta_\theta\sin\delta_\phi\sin\delta_\psi \\ %eq. (31)
  B_2 & = & \cos\delta_\phi\sin\delta_\psi+\cos\delta_\theta\sin\delta_\phi\cos\delta_\psi \\ %eq. (32)
  B_3 & = & \sin\delta_\phi\cos\delta_\psi+\cos\delta_\theta\cos\delta_\phi\sin\delta_\psi \\ %eq. (33)
  B_4 & = & \sin\delta_\phi\sin\delta_\psi-\cos\delta_\theta\cos\delta_\phi\cos\delta_\psi.   %eq. (34)
\end{eqnarray}
It can be verified using an algebraic calculation software that $\theta$ and $\phi$ 
given by equations (29) and (30) satisfy equations (20) and (21) with $c_1 = 0$, 
what means that the latter equations describe correctly the precessional motion around 
an axis oriented arbitrarily in space.

Before proceeding to the solution of equations (20)-(23) let us consider the elasticity 
of the Earth. This was first taken into account by Love \cite{1love} and Larmor \cite{2larmor} because the 
predicted free precession period was of 305 days while the observed value was of around 
435 days. The correction due to Love introduces products of inertia $C_{13}$ and $C_{23}$ 
in Euler equations (8)-(9) for the Earth (see Kaula \cite{12kaula}) which become
\begin{eqnarray}
  && A\dot{\omega}_1+(C-A)\omega_2\omega_3-C_{23}\omega^2_3+\dot{C}_{13}\omega_3=\tau_1 \\ %eq. (35)
  && A\dot{\omega}_2-(C-A)\omega_1\omega_3+C_{13}\omega^2_3+\dot{C}_{23}\omega_3=\tau_2 %eq. (36)
\end{eqnarray}
with
\begin{eqnarray}
  C_{13} &=& \frac{k_2R^5_E\omega_3}{3G}\omega_1        \\   %eq. (37)
  C_{23} &=& \frac{k_2R^5_E\omega_3}{3G}\omega_2     %eq. (38)
\end{eqnarray}
where $R_E$ is the Earth mean radius and $k_2$ is the so called Love number. Therefore 
we obtain as a result
\begin{eqnarray}
 && \left(A+\frac{k_2R^5_E\omega_3^2}{3G}\right)\dot{\omega}_1+
      \left(C-A-\frac{k_2R^5_E\omega_3^2}{3G}\right)\omega_2\omega_3=\tau_1  \\ %eq. (39)
 && \left(A+\frac{k_2R^5_E\omega_3^2}{3G}\right)\dot{\omega}_2-
      \left(C-A-\frac{k_2R^5_E\omega_3^2}{3G}\right)\omega_1\omega_3=\tau_2.  %eq. (40)     
\end{eqnarray}
By comparing this with equations (8)-(9) we see that the overall effect upon the last 
ones by the introduction of the elasticity of the Earth is to promote the transformation
\begin{equation}
  A\rightarrow A+\frac{k_2R^5_E\omega_3^2}{3G}.       %eq. (41)
\end{equation}
The Love number $k_2$ is left as one of the free parameters of this model.

We solve numerically the set of four second-order differential equations given by 
(20)-(23), after performing the transformation given by equation (41), through the 
Runge-Kutta-Fehlberg (RKF) method. The time step upper limit is 1 solar 
day/1000 = 86.4 s; changing this limit to 1 solar day/10000 did not cause meaningful 
modifications in the results. We have 11 free parameters, {\it i. e.}, the Love 
number $k_2$, 
the momenta of inertia $A^*$, $C^*$ of the internal ellipsoid, and the following that 
establish the initial conditions: the components $\omega^*_{1^*}$, $\omega^*_{2^*}$, 
$\omega^*_{3^*}$ (recall that $\omega^*_{3^*}$ is constant) of the angular velocity 
$\boldsymbol{\omega}^*$ in the principal axes 1$^*$2$^*$3$^*$ of the internal 
ellipsoid, the Euler angles $\theta_{12} = \alpha$, $\phi_{12}$, $\psi_{12}$ 
that give the spatial orientation of the 1$^*$2$^*$3$^*$ frame with respect to the 
123 frame, and the angular velocities $\dot{\phi}$  and  $\dot{\phi}^*$. On the other 
hand, from equations (14) and (15) we have
\begin{equation}
  \dot{\theta}=\pm\sqrt{\omega_1^2+\omega_2^2-\dot{\phi}^2\sin^2\theta}  %eq. (42)
\end{equation}
from which we can determine the initial value of $\dot{\theta}$ unless by its sign. 
In an analogous way we have
\begin{equation}
  \dot{\theta}^*=\pm\sqrt{\omega^{*2}_{1^*}+\omega^{*2}_{2^*}-\dot{\phi}^{*2}
          \sin^2\theta^*}.                                        %eq. (43)
\end{equation}
>From equations (42)-(43) we can also find bounds for the initial guesses of 
$\dot{\phi}$ and  $\dot{\phi}^*$
\begin{eqnarray}
  && -\frac{\sqrt{\omega_1^2+\omega_2^2}}{\left|\,\sin\theta\,\right|}\leq\dot{\phi}\leq
       \frac{\sqrt{\omega_1^2+\omega_2^2}}{\left|\,\sin\theta\,\right|} \\ %eq. (44)
  && -\frac{\sqrt{\omega_{1^*}^{*2}+\omega_{2^*}^{*2}}}{\left|\,\sin\theta^*\right|}\leq
     \dot{\phi}^*\leq
     \frac{\sqrt{\omega_{1^*}^{*2}+\omega_{2^*}^{*2}}}{\left|\,\sin\theta^*\right|}. %eq. (45)
\end{eqnarray}
Besides, from energy conservation of the Earth-ellipsoid system, we have the following 
relation to be satisfied by the initial guesses of $\omega_{1^*}^*$ and $\omega_{2^*}^*$
\begin{equation}
  \omega_{1^*}^{*2}+\omega_{2^*}^{*2} \geq \frac{c_1}{A}\left[\cos(2\alpha)-1\right]
    -\frac{A}{A^*}\left( \omega_1^2+\omega_2^2\right)-\frac{2\omega_3}{A^*}
     \left( C_{13}\omega_1 + C_{23}\omega_2\right)                       %eq. (46)
\end{equation}
where $\alpha$ is the initial guess for the angle between the axes 3 and 3$^*$, and 
$\omega_1$ and $\omega_2$ are the initial components of the Earth angular velocity 
$\boldsymbol{\omega}$ in the axes 1 and 2, respectively. 

H\"opfner \cite{7hop,8hop} has managed to filter the Chandler wobble (CW) and the annual wobble 
(AW) as major components from the data related to the Earth polar motion of its rotation 
axis around the polar principal axis. These data have being accumulated for more than 
100 years and a compilation of them is published periodically by Gross \cite{13gross} from NASA-JPL. 
We use a relation between the polar motion, described as the two angles
$\widehat{{\it PMX}}$ and  $\widehat{{\it PMY}}$, and the orientation of the Earth rotation axis, 
described by the angles $\theta_\omega$ and $\phi_\omega$ (see Fig. 5). As usual, the 
1-axis is towards the Greenwich meridian and the 2-axis is towards the 90$^\circ$E 
meridian. The angles $\theta_\omega$ and $\phi_\omega$ can be calculated from
\begin{eqnarray}
  && \tan\theta_\omega=\pm\sqrt{\tan^2\widehat{{\it PMX}}+\tan^2
                               \widehat{{\it PMY}}} \\               %eq. (47)
  && \tan\phi_\omega=\frac{\tan\widehat{{\it PMY}}}{\tan\widehat{{\it PMX}}}. %eq. (48)
\end{eqnarray}
Therefore by knowing $\widehat{{\it PMX}}$ and  $\widehat{{\it PMY}}$ at the start of a 
Chandler wobble, taken from H\"opfner \cite{7hop} calculations, and by knowing the value of the 
length-of-day (LOD) at that moment, taken from Gross \cite{13gross} data, we can calculate the initial 
values of $\omega_1$ and $\omega_2$, which are necessary in equations (42), (44) and (46), 
and the value of $\omega_3$, which is considered as constant during the motion. Furthermore, we can 
calculate then the initial components of the Earth angular momentum {\bf L} in the direction 
of the axes 123 as
\begin{eqnarray}
  L_1 &=& A\omega_1+C_{13}\omega_3         \\ %eq. (49)
  L_2 &=& A\omega_2+C_{23}\omega_3         \\ %eq. (50)
  L_3 &=& C\omega_3+C_{13}\omega_1+C_{23}\omega_2.  %eq. (51)
\end{eqnarray}
Once we have guessed the initial values of $\omega^*_{1^*}$, $\omega^*_{2^*}$, $\omega^*_{3^*}$,  
we can also calculate the initial values of the components of ${\bf L}^*$ in the axes 1*2*3* by
\begin{eqnarray}
  L^*_{1^*} &=& A^*\omega^*_{1^*}    \\  %eq. (52)
  L^*_{2^*} &=& A^*\omega^*_{2^*}    \\  %eq. (53)
  L^*_{3^*} &=& C^*\omega^*_{3^*}.       %eq. (54)
\end{eqnarray}
With these quantities available we can calculate the components of the total angular momentum 
${\bf L}_t$ in the axes 123
\begin{eqnarray}
  && \hspace{-14.5cm}L_{t1} = L_1+L^*_1 = L_1+L^*_{1^*}(\cos\phi_{12}\cos\psi_{12}-
     \cos\alpha\sin\phi_{12}\sin\psi_{12})      \nonumber  \\
     \hspace{3cm}\mbox{}-L^*_{2^*}(\cos\phi_{12}\sin\psi_{12}+\cos\alpha\sin\phi_{12}\cos\psi_{12})+
     L^*_{3^*}\sin\alpha\sin\phi_{12}             \\       %eq. (55)
  && \hspace{-14.5cm}L_{t2} = L_2+L^*_2 = L_2+L^*_{1^*}(\sin\phi_{12}\cos\psi_{12}+
     \cos\alpha\cos\phi_{12}\sin\psi_{12})      \nonumber  \\
     \hspace{3cm}\mbox{}+L^*_{2^*}(-\sin\phi_{12}\sin\psi_{12}+\cos\alpha\cos\phi_{12}\cos\psi_{12})-
     L^*_{3^*}\sin\alpha\cos\phi_{12}              \\      %eq. (56)
  && \hspace{-14.5cm}L_{t3} = L_3+L^*_3 = L_3+L^*_{1^*}\sin\alpha\sin\psi_{12}+
          L^*_{2^*}\sin\alpha\cos\psi_{12}+L^*_{3^*}\cos\alpha.  %eq. (57)
\end{eqnarray}
Here, $\alpha=\theta_{12}$, $\phi_{12}$ and $\psi_{12}$ are the initial guesses for the Euler 
angles that describe the 1$^*$2$^*$3$^*$ reference frame with respect to the 123 frame. Since 
we define the Z-axis of the fixed reference frame in the same direction of ${\bf L}_t$ (see Fig. 
6), the initial value of the Euler angle $\theta$, that is necessary for the solution of equations 
(20)-(23), is obtained from
\begin{equation}
  \cos\theta=\frac{{\bf L}_t\cdot{\bf u}_3}{L_t}=\frac{L_{t3}}{\sqrt{L^2_{t1}+L^2_{t2}+L^2_{t3}}}  %eq. (58)
\end{equation}
where ${\bf u}_3$ is a unit vector in the direction of the 3-axis. In order to calculate 
the initial value of $\phi$ we construct the Y-axis as orthogonal to the ${\bf L}^*{\bf L}$-plane 
(Fig. 6), taking the initial values of such vectors, and the X-axis orthogonal to the YZ-plane, 
as usual. Then
\begin{equation}
  \cos\phi=-\frac{u_{3Y}}{u_{3XY}}=\frac{L_{t2}L_1-L_{t1}L_2}{\left|{\bf L}\times 
    {\bf L}_t \right|\sin\theta}                                              %eq. (59)
\end{equation}
where $u_{3Y}$ and $u_{3XY}$ are the projections of ${\bf u}_3$ in the Y-axis and in the XY-plane, 
respectively. In order to determine $\phi$ completely we have also to express
\begin{equation}
  \sin\phi=\frac{u_{3X}}{u_{3XY}}=\frac{L_{t2}(L_{t2}L_3-L_{t3}L_2)-L_{t1}(L_{t3}L_1-L_{t1}L_3)}
     {L_t\left|{\bf L}\times{\bf L}_t \right|\sin\theta}.                   %eq. (60)
\end{equation}
In an analogous way we have the projections of ${\bf L}_t$ in the axes 1*2*3*
\begin{eqnarray}
  && \hspace{-14.5cm}L_{t1^*} = L^*_{1^*}+L_{1^*} = L^*_{1^*}+L_1(\cos\phi_{12}\cos\psi_{12}-
     \cos\alpha\sin\phi_{12}\sin\psi_{12})      \nonumber  \\
     \hspace{3cm}\mbox{}+L_2(\sin\phi_{12}\cos\psi_{12}+\cos\alpha\cos\phi_{12}\sin\psi_{12})+
     L_3\sin\alpha\sin\psi_{12}             \\       %eq. (61)
  && \hspace{-14.5cm}L_{t2^*} = L^*_{2^*}+L_{2^*} = L^*_{2^*}-L_1(\cos\phi_{12}\sin\psi_{12}+
     \cos\alpha\sin\phi_{12}\cos\psi_{12})      \nonumber  \\
     \hspace{3cm}\mbox{}+L_2(-\sin\phi_{12}\sin\psi_{12}+\cos\alpha\cos\phi_{12}\cos\psi_{12})+
     L_3\sin\alpha\cos\psi_{12}              \\      %eq. (62)
  && \hspace{-14.5cm}L_{t3^*} = L^*_{3^*}+L_{3^*} = L^*_{3^*}+L_1\sin\alpha\sin\phi_{12}-
          L_2\sin\alpha\cos\phi_{12}+L_3\cos\alpha.  %eq. (63)
\end{eqnarray}
and the initial values for the Euler angles $\theta^*$ and $\phi^*$ are obtained from
\begin{eqnarray}
  && \cos\theta^*=\frac{{\bf L}_t\cdot{\bf u}^*_3}{L_t}=\frac{L_{t3^*}}{\sqrt{L^2_{t1^*}+
      L^2_{t2^*}+L^2_{t3^*}}}                       \\  %eq. (64)
  &&   \cos\phi^*=-\frac{u^*_{3Y}}{u^*_{3XY}}=\frac{L_{t1^*}L^*_{2^*}-L_{t2^*}L^*_{1^*}}
      {\left|{\bf L}^*\times{\bf L}_t \right|\sin\theta^*}                 \\  %eq. (65)
  && \sin\phi^*=\frac{u^*_{3X}}{u^*_{3XY}}=\frac{L_{t2^*}(L_{t3^*}L^*_{2^*}-L_{t2^*}L^*_{3^*})
      -L_{t1^*}(L_{t1^*}L^*_{3^*}-L_{t3^*}L^*_{1^*})}
     {L_t\left|{\bf L}^*\times{\bf L}_t \right|\sin\theta^*}.                   %eq. (66)
\end{eqnarray}

While integrating equations (20)-(23) we can calculate the CW motion through (see Fig. 5)
\begin{eqnarray}
  && \tan\widehat{{\it PMX}} = \frac{\omega_1}{\omega_3}    \\ %eq. (67)
  && \tan\widehat{{\it PMY}} = \frac{\omega_2}{\omega_3}       %eq. (68)
\end{eqnarray}
with $\omega_1$ and $\omega_2$ given by equations (14) and (15). The necessary values of 
$\psi$  are obtained step-by-step through numerical integration of equation (16).

\section{\bf Results and discussion}

With the aim of getting 10 unknown parameters of our model (all but $k_2$) we find that 
it is enough to establish the following conditions: 1) reproduce the Chandler component 
of the Earth rotation axis position predicted by H\"opfner \cite{7hop} at the start of a wobble; 
2) minimize the length of the prograde (counter-clockwise) trajectory followed during 
that wobble motion. The second condition is necessary in order to avoid intrusive 
nutations that might emerge accompanying the precessional motion. This means that we 
minimize, for a given initial guess of $k_2$, the sum of: a) the squared distance from the 
model calculed Earth rotation axis position $(\widehat{\it PMX},\widehat{\it PMY})$
at the start of the wobble and the one predicted by H\"opfner; b) the wobble trajectory length 
in the plane $(\widehat{\it PMX},\widehat{\it PMY})$ obtained by numerical integration. For 
this minimization 
process we used the {\tt POWELL} routine from the Numerical Recipes package (Press {\it et al.} 
\cite{14press}). However the generalized simulated annealing ({\tt GSA}) code (Mundim and Tsallis \cite{15mundim}, 
Dall'Igna Junior {\it et al.} \cite{16dall}) has proved to be very effective in a first step run in order 
to prevent   getting stuck in local minima. The 11th parameter, $k_2$, is then found by 
reproducing the Chandler component of the Earth rotation axis position at the end of the 
wobble, with reference to H\"opfner's prediction, so that $k_2$ is responsible for the 
adjustment of a particular wobble period. We show in Figs. 7-9 some of our results for the CW 
in three different epochs, in comparison with the results due to H\"opfner \cite{7hop}, which are 
represented by dots. The first one, for which $\omega^*_{3^*}$ is negative, has the typical quality 
of providing
reasonable reproduction of the motion. The next Figures show the two exceptions to this rule, which
tend to present almost circular trajectories instead of, in those epochs, the slightly elliptical 
ones due to H\"opfner. This may be caused by the above mentioned 
criteria -- mainly condition 2) -- adopted in the fit process. We also show in Table I the 
corresponding values for the 11 parameters (with respective numerical uncertainties shown in 
Table II), while in Table III we have the initial values 
of $\dot{\theta}$ and $\dot{\theta}^*$ calculated from equations (42) and (43) with 
appropriate signs and the corresponding values of the torque strength $c_1$. As expected -- see 
comments after equation (7) -- $\alpha$ turns out to be 
close to 90$^\circ$. On the other hand, parameters $\omega^*_{3^*}$ (the main angular 
velocity component of the ellipsoid), $A^*$, $C^*$ and $k_2$ present oscillations 
when examining the respective columns in the Table I, being more noticeable for the 
44330-44755 interval. For $k_2$, since it is connected to the period and that particular 
wobble has a short one -- see Table IV for a comparison between the presently calculated 
periods and the ones due to H\"opfner which they shall reproduce -- this is not surprising 
although undesirable. H\"opfner \cite{7hop,8hop}, like other authors (see {\it e. g.} Liu {\it et 
al.} \cite{17liu}, Wang \cite{18wang}), found variable CW periods but this is not a clearly solved question 
since some investigators have another opinion (see {\it e. g.} Vicente and Wilson \cite{19vic}, 
Jochmann \cite{20jo}, Liao and Zhou \cite{21liao}, Guo {\it et al.} \cite{22guo}). For the sake of comparison, 
values of $k_2$ obtained by some other authors are 0.284 (Kaula \cite{12kaula}), 0.30088 (model 
1066A of Gilbert and Dziewonski \cite{23gil}) which are not far from what we got with the exception 
of the 44330-44755 case. 

By comparing the values of $C^*$ with momenta of inertia of natural satellites present in the solar 
system we find that they range from about Miranda's (Uranus) to Charon's (Pluto). Since 
we are able only to calculate the DM ellipsoid principal momenta of inertia,
which are related to the ellipsoid mass $M^*$ and their equatorial and polar radii $a^*$ and
$c^*$ through $A^* = M^*(a^{*2}+c^{*2})/5$ and $C^* = 2M^*a^{*2}/5$, we cannot establish the 
ellipsoid mass uniquely, and in consequence, the fraction of DM present in the Earth is not predicted. 
Notice that, for an oblate ellipsoid, we have $c^*\leq a^*$ and, therefore, the relation 
$C^*/2 \leq A^* \leq C^*$ holds and it was required to be satisfied in our numerical procedure. 
Just to have an idea of the order of magnitude of $M^*$, let us suppose that $a^*\approx 2350$ km,
{\it i. e.}, the ellipsoid has the average radius of the outer core of the Earth. From the 
values of $C^*$ in Table I, this implies that the fraction of DM present in the Earth would 
lie in the range $6 \times 10^{-7}$ to $9 \times 10^{-5}$. We emphasize that, 
according to the model, the Earth motion and its gravitational interaction with other 
bodies have already incorporated the presence of DM through what has been known effectively
as ``Earth mass", so that the outcome of its presence might be more easily observed through 
the Earth wobble.

Following the hypothesis of changing CW periods we present in Fig. 10 the correlation 
between period and average amplitude. We show H\"opfner's results \cite{7hop} and ours. The last ones are 
fitted by the full line, which has the form: period (in days) = 440--3101.47$\times$exp[\mbox{--0.0347946} 
$\times$ amplitude (in mas)]. As a reference we also present the fits due to Liu et al \cite{17liu} 
related to the epochs 1912-1928 and 1936-1948. The correlation clearly varies with time.

One might argue that the parameter fluctuations in Table I indicate that this model is 
not sound. However those fluctuations may be caused by the model simplistic -- although 
convenient at this stage -- approach that we are considering the DM ellipsoid 
the only source responsible for the complexity of Chandler wobble, what obviously is not 
true. Therefore, the predicted $A^*$ and $C^*$ dark matter ellipsoid parameters should be 
considered only as upper limit values. 

Another point that could be raised against the model is that it is just emulating one or 
more of the attributed causes for CW -- as listed in the Introduction -- and, in consequence, 
the DM ellipsoid existence is not real. However, it can be shown that externally to the DM 
ellipsoid, whose rotation axis is almost orthogonal to the Earth's, it generates a 
non-isotropic, time dependent, gravitational field with two contributions, with respect to 
a ${\bf u}_1{\bf u}_2{\bf u}_3$ reference frame attached to the Earth. The first term 
is radial:
\begin{equation}
  a_r=\frac{3G(C^*-A^*)}{2r^4}\left[3\sin^2\theta(\cos\phi\sin\omega_Tt+
    \sin\phi\cos\omega_Tt)^2-1\right].                                      %eq. (69)
\end{equation}
Here, $\theta$ is the co-latitude, $\phi$ is the longitude and $\omega_T$ is the Earth
angular velocity. Of course, $\omega_Tt$ is defined except for a phase that would establish 
the position of the ellipsoid rotation axis with respect to the Greenwich meridian (for 
instance, in the direction of ${\bf u}_1$), at a given time. The second contribution is
\begin{eqnarray}
  && \hspace{-1.0cm}{\bf a}_{\theta^*}=\frac{3G(C^*-A^*)}{2r^4}\sin2\theta^*\left[(
      \cos\theta^*\sin\phi^*\cos\omega_Tt-\sin\theta^*\sin\omega_Tt){\bf u}_1 \right. \nonumber \\
  &&  \hspace{2.2cm} \left. \mbox{} -(\cos\theta^*\sin\phi^*\sin\omega_Tt+\sin\theta^*
      \cos\omega_Tt){\bf u}_2+\cos\theta^*\cos\phi^*{\bf u}_3\right]   %eq. (70)
\end{eqnarray}
where
\begin{eqnarray}
  && \sin\theta^*=\sqrt{\cos^2\theta+\sin^2\theta(\cos\phi\cos\omega_Tt-\sin\phi\sin
      \omega_Tt)^2} \\                                           %eq. 71
  && \cos\theta^*=\sin\theta(\cos\phi\sin\omega_Tt+\sin\phi\cos\omega_Tt)) \\ %eq. (72)
  && \sin\phi^*=\frac{\sin\theta(\cos\phi\cos\omega_Tt-\sin\phi\sin\omega_Tt)}
     {\sqrt{\cos^2\theta+\sin^2\theta(\cos\phi\cos\omega_Tt-\sin\phi\sin\omega_Tt)^2}} \\ %eq. (73)
  && \cos\phi^*=\frac{\cos\theta}{\sqrt{\cos^2\theta+\sin^2\theta(\cos\phi\cos\omega_Tt-
     \sin\phi\sin\omega_Tt)^2}}.
\end{eqnarray}
Note that, in the Earth's reference frame, the DM ellipsoid gives a complete turn in a 
day. Supposing the DM ellipsoid is smaller than the Earth core, this gravitational field 
certainly enforces viscous flow in the outer core, which is in a fluid state. As a first 
consequence, it can provide the necessary energy to the magneto-hydrodynamic motion in 
the outer core, which is closely related to the generation of the geomagnetic field. 
This could give a plausible explanation to one of the basic problems of geophysics 
(see Stacey, ref. \cite{9stacey}), namely, what is the source of this energy. However 
the numerical solution of this motion through the Navier-Stokes and energy conservation 
equations suffers lack of knowledge of the flowing material density dependence with 
respect to pressure and temperature. Secondly, it can generate heat in the core, that 
would contribute to decrease the shortfall of 0.7 TW to the energy necessary to 
maintain the adiabatic temperature gradient in the core, leaving more power to drive the 
geomagnetic field dynamo \cite{9stacey}. On the other hand, Mack {\it et al.} 
\cite{mack} have concluded that DM is unlikely to contribute not only to Earth's internal 
heat flow but also to hot-Jupiter exoplanets. However they studied the contribution of DM self-annihilation to heat generation, what is a different idea from ours. Conversely, 
Adler \cite{adl} has also studied the contribution of self-annihilating and 
non-self-annihilating DM accretion to the internal heat of the Earth, Jovian planets 
and hot-Jupiter exoplanets. His conclusion is that this process is plausible provided 
efficient DM capture is occurring. In our description, it is feasible to imagine DM 
ellipsoids also present is such bodies, generating heat by enforcing internal viscous 
flow, like in Earth. Therefore this model can mean much more than mere emulation of known 
suggested causes for CW but it also touches open questions like geomagnetic field dynamo 
and heat generation in the Earth's outer core and in other planets.

Rather than giving the final answer for the problem 
that has raised the attention of geophysicists and astronomers for several decades, we 
expect that this calculation may open room for a new and potentially important key component 
in CW which was completely unsuspected untill now, to be considered in more sophisticated 
models. Future development will decide about the real significance of this approach that might
have relevant consequences in our comprehension about dark matter. At least, it has been 
demonstrated that two interacting ellipsoids can have Chandler-like wobble.

\vspace {1cm}
\noindent {\bf ACKNOWLEDGMENTS}
\vspace {3mm}

The author would like to thank Prof. Joachim H\"opfner (retired from GFZ-Potsdam) for 
enlightening correspondence and for providing unpublished results, Prof. Kleber C. Mundim, 
from Chemistry Institute-University of Brasilia, for allowing the use of his {\tt GSA} code, Prof. 
Nivaldo A. Lemos, from Universidade Federal Fluminense, for correspondence in the early stages 
of this work, Prof. Marcos D. Maia, from Institute of Physics-University of Brasilia, for 
his interest and for pointing me some related literature, Prof. Marcus B. Lacerda Santos, 
also from Institute of Physics-University of Brasilia, for stimulating discussions, and 
MEC-SESu for tutorship (PET program). Finally, thanks are due to the Referee of this paper, 
for interesting remarks.

\pagebreak

\pagebreak
\noindent {\bf FIGURE CAPTIONS} \vspace{7mm} \\
\noindent {\bf Figure 1} --
Two ellipsoids tilted by an angle $\alpha$ around the common equatorial principal 
axes 1 and 1$^*$. The spherical coordinates ($r$, $\theta$, $\phi$) that give the 
position of an infinitesimal mass $dm$ with respect to the 1$^*$2$^*$3$^*$ reference frame 
are also shown.

\vspace{7mm}

\noindent {\bf Figure 2} --
Definition of the Euler angles $\theta$, $\phi$, $\psi$.

\vspace{7mm}

\noindent {\bf Figure 3} --
Space and body cones generated by the Eulerian (free) precession. The axes 1, 2 
(equatorial principal axes) and 3 (polar principal axis) are fixed in the oblate 
ellipsoid and follow its motion. The angular momentum {\bf L} defines a fixed 
direction in space while the angular velocity $\boldsymbol{\omega}$ gives the 
instantaneous rotation axis.

\vspace{7mm}

\noindent {\bf Figure 4} --
Free precession around an arbitrary axis z. The xyz-frame is oriented with respect 
to the XYZ-frame, which is fixed in space, through the Euler angles $\delta_\theta$,
$\delta_\phi$, $\delta_\psi$.

\vspace{7mm}

\noindent {\bf Figure 5} --
Definition of the angles $\widehat{{\it PMX}}$ and $\widehat{{\it PMY}}$ which 
describe the Chandler wobble motion.  The origin of the 123-frame is at the Earth 
center. Axes 1 and 2 are towards the Greenwich meridian and 90$^\circ$E meridian, 
respectively. The 3-axis is the polar principal axis. The angular velocity 
$\boldsymbol{\omega}$ is also shown as well as its spherical coordinates 
$\theta_\omega$ and $\phi_\omega$.

\vspace{7mm}

\noindent {\bf Figure 6} --
Definition of the frame XYZ fixed in space which is convenient to describe both 
interacting ellipsoids motion. The Z-axis coincides with the direction of the 
total angular momentum ${\bf L}_t$. The XZ-plane is the same plane formed by the 
{\bf L} (Earth) and ${\bf L}^*$ (ellipsoid) angular momenta. The unit vectors 
${\bf u}_3$ and ${\bf u}_3^*$, which are in the direction of the polar principal 
axes 3 and 3*, are also shown as well as their spherical coordinates $\theta$, 
$\phi_s$, $\theta^*$, $\phi_s^*$.

\vspace{7mm}

\noindent {\bf Figure 7} --
Calculated (line) and predicted (dots) Chandler wobble prograde (counter-clockwise) 
motion in the interval 44330-44755 (in modified Julian date - MJD) or 
01/APR/ 1980-31/MAY/1981. The asterisk shows the position of the motion start point.
$\widehat{{\it PMX}}$ and $\widehat{{\it PMY}}$ are given in milli-arcsec (mas).

\vspace{7mm}

\noindent {\bf Figure 8} --
Same as in Fig. 7, for the interval 44970-45403 (MJD) or 01/JAN/1982-10/MAR/1983.

\vspace{7mm}

\noindent {\bf Figure 9} --
Same as in Fig. 7, for the interval 50170-50604 (MJD) or 28/MAR/1996-05/JUN/1997.

\vspace{7mm}

\noindent {\bf Figure 10} --
Relation between CW period (in days) and average amplitude (in mas). Circles represent 
results due to H\"opfner \cite{7hop} while triangles show our results. The full line fits the
last points. As a reference, dashed and dotted lines represent fits for the epochs 1912-1928 
and 1936-1948 respectively.

\vspace{7mm}

\noindent {\bf TABLE CAPTIONS} \vspace{7mm} \\
\noindent {\bf Table I} --
Results for the 11 unknown parameters in several epochs: initial values for the 
angular velocities $\dot{\phi}$  and  $\dot{\phi}^*$ (in rad/s), for the components 
$\omega^*_{1^*}$, $\omega^*_{2^*}$, $\omega^*_{3^*}$ (in rad/s) of angular velocity 
$\boldsymbol{\omega}^*$, for the Euler angles $\alpha=\theta_{12}$, $\phi_{12}$, 
$\psi_{12}$ (in degrees); values of the internal ellipsoid principal momenta of 
inertia $A^*$, $C^*$ (in ${\rm kg}\cdot{\rm m}^2$) and Love number $k_2$.

\vspace{7mm}

\noindent {\bf Table II} --
Numerical uncertainty estimates (in \%) of the parameters shown in Table I.

\vspace{7mm}

\noindent {\bf Table III} --
Calculated initial values of $\dot{\theta}$ and $\dot{\theta}^*$ (in rad/s), with 
respective appropriate signs, and torque strength $c_1$ (in ${\rm N}\cdot{\rm m}^2$).

\vspace{7mm}

\noindent {\bf Table IV} --
CW periods resulting from the model in comparison with the values from H\"opfner 
\cite{7hop} (in days).

\pagebreak             
\begin{sidewaystable}
\begin{center}{\bf Table I} \end{center}
%\vspace{0.5 cm}
\begin{center}
\begin{tabular}{crrrrrrrrrrrr}  \hline \\
interval & $\dot{\phi}$\rule{6mm}{0mm} & $\dot{\phi}^*$\rule{1em}{0mm} & $\omega^*_{1^*}$\rule{1em}{0mm}  
& $\omega^*_{2^*}$\rule{1em}{0mm} & $\omega^*_{3^*}$\rule{1em}{0mm} & $\phi_{12}$\rule{1em}{0mm} 
   & $\psi_{12}$\rule{1em}{0mm} & $\alpha$\rule{1em}{0mm} & $A^*$\rule{5.5mm}{0mm} 
   & $C^*$\rule{5.5mm}{0mm} & $k_2$\rule{1em}{0mm}  \\[2mm]
				  \hline \\[2mm]
{\footnotesize 44330-44755} & {\footnotesize   1.386E--7} & {\footnotesize   3.087E--20} 
 & {\footnotesize  9.907E--12} & {\footnotesize  4.140E--10} & {\footnotesize --1.592E--4} 
 & {\footnotesize 358.849} & {\footnotesize 353.219} & {\footnotesize 90.001} & {\footnotesize 6.334E+32} 
 & {\footnotesize 1.264E+33} & {\footnotesize 0.2672} \\
{\footnotesize 44970-45403} & {\footnotesize --2.704E--5} & {\footnotesize --9.685E--7} 
 & {\footnotesize --9.789E--7} & {\footnotesize --2.088E--9} & {\footnotesize 4.952E--4} 
 & {\footnotesize 173.698} & {\footnotesize 337.477} & {\footnotesize 89.911} & {\footnotesize 2.399E+31} 
 & {\footnotesize 3.334E+31} & {\footnotesize 0.2801} \\
{\footnotesize 45700-46135} & {\footnotesize   5.711E--5} & {\footnotesize --4.203E--6} 
 & {\footnotesize --4.262E--6} & {\footnotesize --2.128E--9} & {\footnotesize 6.853E--4} 
 & {\footnotesize 359.988} & {\footnotesize 359.999} & {\footnotesize 89.847} & {\footnotesize 1.079E+31} 
 & {\footnotesize 1.125E+31} & {\footnotesize 0.2835} \\
{\footnotesize 46340-46773} & {\footnotesize --7.419E--5} & {\footnotesize   1.823E--6} 
 & {\footnotesize   3.935E--6} & {\footnotesize --2.410E--9} & {\footnotesize 1.089E--3} 
 & {\footnotesize 187.700} & {\footnotesize 350.276} & {\footnotesize 89.849} & {\footnotesize 7.429E+30} 
 & {\footnotesize 7.772E+30} & {\footnotesize 0.2815} \\
{\footnotesize 46977-47410} & {\footnotesize   7.205E--5} & {\footnotesize --4.490E--6} 
 & {\footnotesize --4.542E--6} & {\footnotesize --2.128E--9} & {\footnotesize 7.455E--4} 
 & {\footnotesize 359.988} & {\footnotesize 359.999} & {\footnotesize 89.857} & {\footnotesize 1.168E+31} 
 & {\footnotesize 1.211E+31} & {\footnotesize 0.2815} \\
{\footnotesize 47640-48075} & {\footnotesize --3.769E--5} & {\footnotesize  2.098E--6} 
 & {\footnotesize   3.572E--6} & {\footnotesize --2.410E--9} & {\footnotesize 7.727E--4} 
 & {\footnotesize 109.191} & {\footnotesize 353.481} & {\footnotesize 89.832} & {\footnotesize 7.280E+30} 
 & {\footnotesize 7.548E+30} & {\footnotesize 0.2842} \\
{\footnotesize 48260-48696} & {\footnotesize   5.317E--6} & {\footnotesize --1.861E--7} 
 & {\footnotesize --1.862E--7} & {\footnotesize --2.128E--9} & {\footnotesize 3.338E--3} 
 & {\footnotesize 222.469} & {\footnotesize 359.999} & {\footnotesize 90.047} & {\footnotesize 1.947E+31} 
 & {\footnotesize 2.125E+31} & {\footnotesize 0.2861} \\
{\footnotesize 48900-49336} & {\footnotesize --2.505E--5} & {\footnotesize --1.142E--6} 
 & {\footnotesize --1.142E--6} & {\footnotesize  --2.088E--9} & {\footnotesize 5.632E--4} 
 & {\footnotesize 346.867} & {\footnotesize 332.210} & {\footnotesize 89.841} & {\footnotesize 1.731E+31} 
 & {\footnotesize 1.903E+31} & {\footnotesize 0.2869} \\
{\footnotesize 49630-50066} & {\footnotesize   4.133E--5} & {\footnotesize --6.759E--6} 
 & {\footnotesize --6.759E--6} & {\footnotesize --2.128E--9} & {\footnotesize 7.390E--4} 
 & {\footnotesize 359.988} & {\footnotesize 359.999} & {\footnotesize 89.873} & {\footnotesize 9.482E+30} 
 & {\footnotesize 9.710E+30} & {\footnotesize 0.2869} \\
{\footnotesize 50170-50604} & {\footnotesize --3.317E--5} & {\footnotesize --6.301E--7} 
 & {\footnotesize --6.301E--7} & {\footnotesize --2.088E--9} & {\footnotesize 5.230E--4} 
 & {\footnotesize 247.397} & {\footnotesize 333.076} & {\footnotesize 89.877} & {\footnotesize 8.287E+30} 
 & {\footnotesize 9.587E+30} & {\footnotesize 0.2824} \\
			\\[2mm]  \hline
\end{tabular}\end{center}
\end{sidewaystable}

\pagebreak
\begin{sidewaystable}
\begin{center}{\bf Table II} \end{center}
%\vspace{0.5 cm}
\begin{center}
\begin{tabular}{cccccccccccc}  \hline \\
interval & $\left|\frac{\Delta\dot{\phi}}{\dot{\phi}}\right|$\rule{1em}{0mm} & 
   $\left|\frac{\Delta\dot{\phi}^*}{\dot{\phi}^*}\right|$\rule{1.5mm}{0mm} &
   $\left|\frac{\Delta\omega^*_{1^*}}{\omega^*_{1^*}}\right|$\rule{1mm}{0mm} &
   $\left|\frac{\Delta\omega^*_{2^*}}{\omega^*_{2^*}}\right|$\rule{1mm}{0mm} &  
   $\left|\frac{\Delta\omega^*_{3^*}}{\omega^*_{3^*}}\right|$\rule{1mm}{0mm} & 
   $\left|\frac{\Delta\phi_{12}}{\phi_{12}}\right|$\rule{1mm}{0mm} &
   $\left|\frac{\Delta\psi_{12}}{\psi_{12}}\right|$\rule{1mm}{0mm} & 
   $\left|\frac{\Delta\alpha}{\alpha}\right|$\rule{2.5mm}{0mm} & 
   $\left|\frac{\Delta A^*}{A^*}\right|$\rule{2.5mm}{0mm} &
   $\left|\frac{\Delta C^*}{C^*}\right|$\rule{2.5mm}{0mm} & 
   $\left|\frac{\Delta k_2}{k_2}\right|$\rule{0.5mm}{0mm}  \\[2mm]
				  \hline \\[2mm]
{\footnotesize 44330-44755} & {\footnotesize 0.10} & {\footnotesize 4.6E--6} 
 & {\footnotesize 1.3E--7} & {\footnotesize 2.3E--9} & {\footnotesize 71} 
 & {\footnotesize 1.6E--3}  & {\footnotesize 1.6E--2} & {\footnotesize 6.6E--4} 
 & {\footnotesize 1.1E--3} & {\footnotesize 129}  
 & {\footnotesize 1.1E--4} \\
{\footnotesize 44970-45403} & {\footnotesize 1.8E--6} & {\footnotesize 0.95} 
 & {\footnotesize 0.77} & {\footnotesize 3.0E--5} & {\footnotesize 2.5E--7} 
 & {\footnotesize 2.2} & {\footnotesize 0.37} & {\footnotesize 5.0E--2} 
 & {\footnotesize 4.1E--6} & {\footnotesize 28}  
 & {\footnotesize 1.1E--3} \\
{\footnotesize 45700-46135} & {\footnotesize 9.6E--6} & {\footnotesize 1.5} 
 & {\footnotesize 0.16} & {\footnotesize 1.8E--6} & {\footnotesize 1.6E--6} 
 & {\footnotesize 4.5E--8} & {\footnotesize 1.9E--10} & {\footnotesize 9.1E--6} 
 & {\footnotesize 4.3} & {\footnotesize 1.8}  
 & {\footnotesize 1.2E--5} \\
{\footnotesize 46340-46773} & {\footnotesize 3.0E--4} & {\footnotesize 0.44} 
 & {\footnotesize 0.11} & {\footnotesize 3.7E--5} & {\footnotesize 13} 
 & {\footnotesize 16} & {\footnotesize 2.8E--2} & {\footnotesize 3.1E--4} 
 & {\footnotesize 4.6} & {\footnotesize 7.0}  
 & {\footnotesize 6.6E--5} \\
{\footnotesize 46977-47410} & {\footnotesize 1.2E--6} & {\footnotesize 1.3} 
 & {\footnotesize 0.64} & {\footnotesize 2.7E--5} & {\footnotesize 6.2E--7} 
 & {\footnotesize 7.2E--8} & {\footnotesize $<$1E--13} & {\footnotesize 7.9E--6} 
 & {\footnotesize 3.8} & {\footnotesize 3.5}  
 & {\footnotesize 5.8E--5} \\
{\footnotesize 47640-48075} & {\footnotesize 3.6E--4} & {\footnotesize 3.8} 
 & {\footnotesize 0.26} & {\footnotesize 2.0E--5} & {\footnotesize 1.6E--5} 
 & {\footnotesize 6.2} & {\footnotesize 3.5E--2} & {\footnotesize 7.4E--5} 
 & {\footnotesize 4.1} & {\footnotesize 1.3}  
 & {\footnotesize 1.2E--4} \\
{\footnotesize 48260-48696} & {\footnotesize 3.3E--5} & {\footnotesize 2.8E--2} 
 & {\footnotesize 1.7E--3} & {\footnotesize 3.4E--5} & {\footnotesize 38} 
 & {\footnotesize 26} & {\footnotesize 7.7E--10} & {\footnotesize 7.0E--6} 
 & {\footnotesize 14} & {\footnotesize 22}  
 & {\footnotesize 4.9E--5} \\
{\footnotesize 48900-49336} & {\footnotesize 8.3E--5} & {\footnotesize 1.1} 
 & {\footnotesize 1.9} & {\footnotesize 5.0E-5} & {\footnotesize 4.6} 
 & {\footnotesize 0.34} & {\footnotesize 1.4} & {\footnotesize 1.5E--5} 
 & {\footnotesize 9.9} & {\footnotesize 13}  
 & {\footnotesize 1.6E--4} \\
{\footnotesize 49630-50066} & {\footnotesize $<$1E--13} & {\footnotesize 2.6} 
 & {\footnotesize 6.8E--2} & {\footnotesize 2.6E--7} & {\footnotesize 1.6E--3} 
 & {\footnotesize 7.9E--14} & {\footnotesize $<$1E--13} & {\footnotesize 3.2E--9} 
 & {\footnotesize 2.7} & {\footnotesize 0.38}  
 & {\footnotesize 2.5E--5} \\
{\footnotesize 50170-50604} & {\footnotesize 2.7E--5} & {\footnotesize 2.3} 
 & {\footnotesize 1.0E--6} & {\footnotesize 3.0E--6} & {\footnotesize $<$1E--13} 
 & {\footnotesize 1.1E--2} & {\footnotesize 2.2E--2} & {\footnotesize 2.7E--9} 
 & {\footnotesize 18} & {\footnotesize 11}  
 & {\footnotesize 1.3E--4} \\
			\\[2mm]  \hline
\end{tabular}\end{center}
\end{sidewaystable}

\pagebreak
\begin{center}{\bf Table III} \end{center}
\vspace{0.5 cm}
\begin{center}
\begin{tabular}{crrr}  \hline \\
interval & $\dot{\theta}$\rule{2em}{0mm} & $\dot{\theta}$\rule{2em}{0mm} & $c_1$\rule{2em}{0mm} \\[2mm]
				  \hline \\[2mm]
44330-44755  &   2.426E--11  &   4.141E--10 & 9.226E+23 \\
44970-45403  & --2.684E--11  &   1.419E--07 & 1.367E+22 \\
45700-46135  & --4.176E--11  &   7.080E--07 & 6.790E+20 \\
46340-46773  &   3.218E--11  &   3.487E--06 & 5.006E+20 \\
46977-47410  & --2.231E--11  &   6.839E--07 & 6.349E+20 \\
47640-48075  &   3.367E--11  &   2.891E--06 & 3.922E+20 \\
48260-48696  & --4.771E--12  & --4.521E--09 & 2.607E+21 \\
48900-49336  & --9.816E--12  &   1.147E--08 & 2.517E+21 \\
49630-50066  & --5.563E--11  &   3.185E--10 & 3.331E+20 \\
50170-50604  & --4.110E--11  &   4.946E--10 & 1.900E+21 \\
			\\[2mm]  \hline
\end{tabular}\end{center}
\vspace{1cm}

\begin{center}{\bf Table IV} \end{center}
\vspace{0.5 cm}
\begin{center}
\begin{tabular}{ccc}  \hline \\
interval & calculated period & H\"opfner's period \\[2mm]
				  \hline \\[2mm]
44330-44755  & 424.578  &   424.76  \\
44970-45403  & 432.454  &   432.45  \\
45700-46135  & 434.677  &   434.68  \\
46340-46773  & 433.366  &   433.37  \\
46977-47410  & 433.331  &   433.33  \\
47640-48075  & 435.123  &   435.12  \\
48260-48696  & 436.414  &   436.41  \\
48900-49336  & 436.943  &   436.93  \\
49630-50066  & 436.924  &   436.92  \\
50170-50604  & 433.920  &   433.92  \\
			\\[2mm]  \hline
\end{tabular}\end{center}

\end{document}